\documentclass[aps,prb,twocolumn,showpacs,epsfig]{revtex4}
\usepackage{amsmath}
\usepackage{graphicx}
\usepackage{amssymb}

\begin{document}

\title{The planar pyrochlore antiferromagnet: A large-$N$ analysis}
\author{Jean-S\'ebastien Bernier$^{1}$}
\author{Chung-Hou Chung$^{1}$}
\author{Yong Baek Kim$^{1}$}
\author{Subir Sachdev$^{2}$}
\affiliation{$^{1}$Department of Physics, University of Toronto, Toronto,
Ontario, Canada M5S 1A7 \\
$^{2}$Department of Physics, Yale University, P. O. Box 208120, New Haven,
Connecticut 06520-8120}

\date{\today}

\begin{abstract}

We study possible quantum phases of the Heisenberg antiferromagnet on
the planar pyrochlore lattice, also known as the checkerboard lattice
or the square lattice with crossings. It is assumed that the exchange
coupling on the square-lattice-links is not necessarily the same
as those along the crossing links. When all the couplings are the same,
this model may be regarded as a two dimensional analog of the pyrochlore
antiferromagnet. The large-$N$ limit of the Sp($N$) generalized
model is considered and the phase diagram is obtained by analyzing
the fluctuation effects about the $N \rightarrow \infty$ limit.
We find a topologically-ordered $Z_2$-spin-liquid phase in a narrow
region of the phase diagram as well as the plaquette-ordered
and staggered spin-Peierls phases as possible quantum-disordered
paramagnetic phases.

\end{abstract}


\maketitle

\section{Introduction}

Recently frustrated magnets have attracted much interest of theorists
and experimentalists as new kinds of frustrated magnetic systems
have become available for experimental
studies.\cite{ramirez96,kakeyama99,coldea01} Armed with much
progress in understanding the classical frustrated
magnets,\cite{moessner01,moessner98} the
attention is now focused on quantum counterparts. Frustration
often leads to a large degeneracy of the classical ground states
and the resulting frustration-enhanced fluctuations can suppress
classical long-range spin order, encouraging the possibility
of spin-disordered ground states even in two or
three dimensions.\cite{villain,harris91}

It is not obvious, however, what kinds of quantum ground states
may arise as a result of the fluctuations, especially for the
small values of spin. Indeed geometrically frustrated
magnets seem to allow multifarious quantum-disordered
paramagnetic phases, including
various translational-symmetry-breaking phases\cite{sachdev92}
and the quantum spin liquid phases with fractionalized
excitations.\cite{sachdev92,chung01,marston01,chung03}
It has been suggested that some of these quantum-disordered phases
may also arise in underdoped cuprate superconductors, where the frustration
of the spins may occur due to the motion of doped
holes.\cite{vojta00,anderson87}

The possible existence of quantum-disordered phases on the pyrochlore
lattice is a particularly interesting question in view of the fact that
it is generally harder to suppress a long-range spin order in three
dimensions. The progress in understanding this issue has been slower
than the cases of two dimensional frustrated magnets mainly due to the
significantly bigger Hilbert space.\cite{koga00,canals00,tsune01}
To circumvent this difficulty,
the planar version of the pyrochlore lattice, also known as
the checkerboard lattice or the square lattice with crossings,
has been
introduced.\cite{lieb99,palmer01,sondhi01,fouet01,starykh02,sindzingre02,berg03,tcherny03, canals, brenig}
When all the Heisenberg exchange
couplings are the same, the checkerboard lattice has the
same local structure of the pyrochlore lattice; a network of
corner-sharing tetrahedra. Moreover, the size and topology of
the ground state manifold are also identical in two cases.\cite{moessner98}
Therefore one may expect that the studies
of the checkerboard lattice will shed some light on the
pyrochlore lattice problem.

\begin{figure}
\includegraphics[width=6cm]{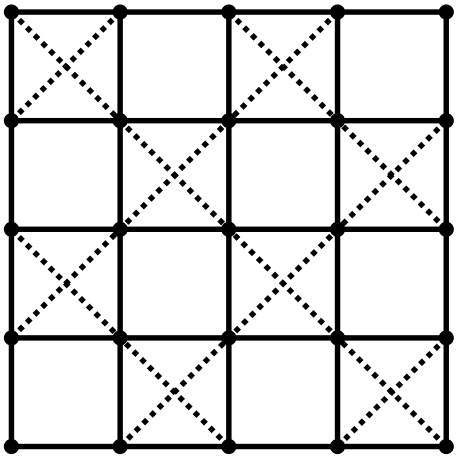}
\caption{\label{fig:cbla} The checkerboard lattice. The exchange $J_1$ acts
between sites separated by the horizontal or vertical links while the
exchange $J_2$ acts across the diagonal links.}
\end{figure}

In this paper, we examine a Heisenberg model on the
checkerboard lattice (illustrated in Fig.\ref{fig:cbla}), where the
Heisenberg exchange couplings on the horizontal and vertical links
are in general different from those along the diagonal links.
The motivation for studying such
a model partly comes from the fact that no symmetry of the lattice can
turn the first neighbor bonds to the second neighbor bonds
so that not all the bonds on its tetrahedra are equivalent
in contrast to the pyrochlore lattice case.
The Hamiltonian for this model can be written as

\begin{equation}
H = J_1 \sum_{\left < ij \right >} \mathbf{S}_i \cdot \mathbf{S}_j
+ J_2 \sum_{\rm diagonals} \mathbf{S}_i \cdot \mathbf{S}_j ,
\label{hamil}
\end{equation}
where $\mathbf{S}_i$ are $S=1/2$ operators at the site $i$.
Here $J_1 > 0$ and $J_2 > 0$ are the antiferromagnetic exchange couplings
on the nearest-neighbor (horizontal and vertical) links and the
second-neighbor (diagonal) links, respectively.

We consider the generalization of the physical spin SU($2$) $\cong$ Sp($1$)
symmetry to Sp($N$) and study the quantum ground states of the
Sp($N$) Hamiltonian in the large-$N$ limit.\cite{sachdev92}
Some of the phases obtained in the large-$N$ limit may not appear in
the SU(2) model. Even these phases, however, may still be relevant to
some physical systems whose microscopic Hamiltonians are ``near'' the
parameter space of the original Hamiltonian. For example, these phases
may arise when the original system is deformed by substitutional doping
or hydrostatic pressure.

More specifically, the first step toward the Sp($N$) generalization starts
with the bosonic representation of the SU(2) spin operators,
$\mathbf{S}_i = \frac{1}{2}b_i^{\dagger \alpha}
\mathbf{\sigma}_\alpha^{\beta}b_{i\beta}$,
where $\alpha, \beta = \uparrow, \downarrow$ labels two possible spin
states of each boson, $b_{i \alpha}$, and the constraint
$n_b = b_i^{\dagger \alpha} b_{i\alpha} = 2S$
must be imposed at each site. Then, apart from an additive constant,
the Heisenberg Hamiltonian can be written as

\begin{equation}
H = -\frac{1}{2}\sum_{ij}J_{ij}(\epsilon_{\alpha \beta}
b_i^{\dagger \alpha}b_j^{\dagger \beta})
(\epsilon^{\gamma \delta} b_{i\gamma} b_{j\delta})
\label{newhamil}
\end{equation}
where $J_{ij} = J_1$ on the horizontal and vertical links, and
$J_{ij} = J_2$ on the diagonal links. Here $\epsilon_{\alpha \beta}$
is the antisymmetric tensor of SU(2).
The generalization to Sp($N$) symmetry can be achieved by introducing
$N$ flavors of bosons on each site. The constraint must be modified to
$n_b = b_i^{\dagger \alpha} b_{i\alpha} = 2NS$, where
$\alpha = 1, ..., 2N$ is a Sp($N$) index. For the physical case, $N=1$,
$S$ takes half-integer values.
The corresponding Sp($N$) Hamiltonian is
\begin{equation}
H = -\frac{1}{2N}\sum_{ij}J_{ij}(\mathcal{J}_{\alpha \beta}b_i^{\dagger \alpha}
b_j^{\dagger \beta})(\mathcal{J}^{\gamma \delta} b_{i\gamma} b_{j\delta}),
\label{spnhamil}
\end{equation}
where $\mathcal{J}^{\alpha \beta} = \mathcal{J}_{\alpha \beta} =
- \mathcal{J}_{\beta \alpha}$ is the generalization of the $\epsilon$ tensor
of SU(2); it is a $2N \times 2N$ matrix that contains $N$ copies of $\epsilon$
along its center block diagonal and vanishes elsewhere.\cite{sachdev92}

In the $N \rightarrow \infty$ limit at a fixed boson density per flavor,
$n_b/N = 2S = \kappa$, a mean field theory for $S = \kappa/2$ can be
obtained and the fluctuations about the mean field state give rise to
a gauge theory.\cite{sachdev92}
The mean-field phase diagram at $N \rightarrow \infty$ is shown
in Fig.\ref{fig:pd} as a function of $J_2/(J_1+J_2)$ and $1/S$.
At large values of $S$, various magnetically long-range-ordered (LRO)
phases appear and are represented by the ordering wavevector $(q_1,q_2)$
measured in units of 1/(nearest-neighbor spacing).
The short-range-ordered (SRO) phases at small values of $S$
correspond to the quantum-disordered phases with
short-range equal-time spin correlations enhanced at the
corresponding wavevectors.

One of the most notable findings in our studies is the existence
of a quantum-spin-liquid phase with
$Z_2$-fractionalized excitations\cite{z2}
in a narrow region of the phase diagram (denoted by $(\pi,q)$ SRO).
This region, $J_1 \not= J_2$ and the small values of $S$,
has not been explored in previous works.
After taking into account fluctuation effects, the plaquette-ordered
state emerges from the $(\pi,\pi)$ SRO phase; similarly a staggered
spin-Peierls state arises from the $(\pi,0)$ SRO phase.
The result on the $(\pi,\pi)$ SRO phase is in accordance
with the previous finding of the plaquette-ordered phase for
$J_1=J_2$ and small values of $S$.\cite{palmer01,sondhi01,fouet01}
The transition from the $Z_2$ quantum-spin-liquid to the
plaquette-ordered phase or the staggered spin-Peierls phase is
described by a $Z_2$ gauge theory.\cite{z2,balents01,nayak01,demler02}
Our results may be also consistent with the large-$S$ study
of the SU(2) model with $J_1 \not= J_2$.\cite{tcherny03}
Detailed comparison with the previous works will be made later
in this paper.

The rest of this paper is organized as follows.
In section II, we describe the mean-field phase diagram and
various mean-field phases at $N \rightarrow \infty$. In section III,
the effect of singular fluctuations at finite $N$ on the paramagnetic
phases is discussed. We explain the relation between the
previous works and ours, and conclude in section IV.

\begin{figure}
\includegraphics[width=5cm]{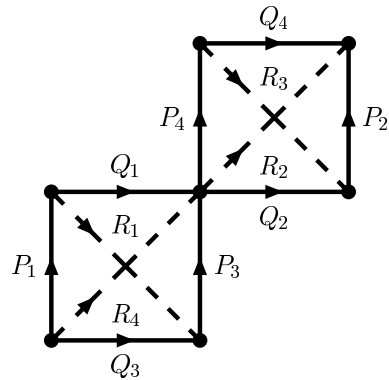}
\caption{\label{fig:cuc} The four-site unit cell of the checkerboard lattice
and the twelve link variables $Q_{ij}$.}
\end{figure}

\section{Mean Field Phase Diagram}

\begin{figure}
\includegraphics[height=6cm,width=8.5cm]{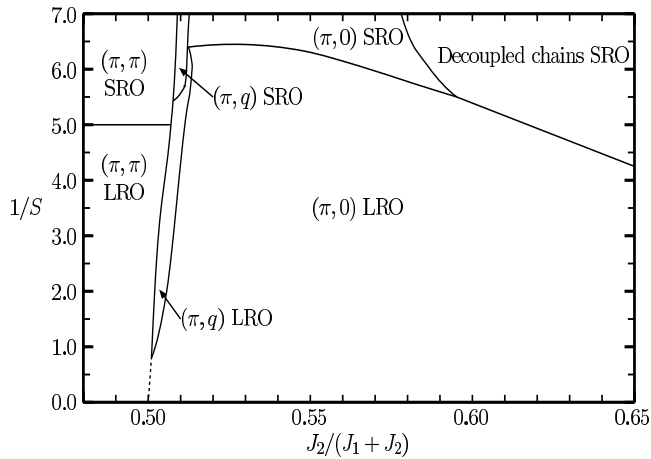}
\caption{\label{fig:pd} [Dashed line: first order transition. Solid line:
continuous transition.] Large-$N$ phase diagram of the Sp($N$) checkerboard
lattice model as a function of $J_2/(J_1+J_2)$ and $1/S$. The LRO phases break
spin-rotation symmetry. The spin order is collinear and commensurate in
the $(\pi,\pi)$ and $(\pi,0)$ LRO phases while it is helical and incommensurate in
the $(\pi,q)$ LRO phase. The SRO phases preserve spin-rotation invariance.
At $J_1/J_2=1$, we find zero expectation value of the diagonal bond variables
at all $\kappa=2S$. The long-range N\'eel order develops above $\kappa=2S\approx0.4$;
below this value, the N\'eel correlations are only short-ranged.
In addition, for large values of $S$, we get the N\'eel $(\pi,\pi)$ LRO for
$J_2 < J_1$ and $(\pi,0)$ LRO for $J_2 > J_1$.}
\end{figure}

The mean-field theory can be obtained by introducing the directed link fields,
$Q_{ij}=-Q_{ji}$. These fields are used to decouple the quartic boson interactions
in $\mathcal{S}$ by a Hubbard-Stratonovitch transformation. After this decoupling,
the effective action contains the terms
\begin{equation}
\mathcal{S}=\int{d\tau \sum_{i>j}\frac{J_{ij}}{2}\left[N \lvert Q_{ij}\lvert^2-Q_{ij}
\mathcal{J}_{\alpha\beta}b_i^{\alpha}b_j^{\beta}+c.c.\right]+\cdots},
\label{action}
\end{equation}
where $\tau$ is the imaginary time and the ellipses represent standard terms which
impose the canonical boson commutation relations and the constraint.\cite{sachdev92}
At the saddle point of the action, we get
\begin{equation}
\left < Q_{ij} \right > = \frac{1}{N} \left < \mathcal{J}^{\alpha\beta}
b_{i\alpha}^{\dagger}b_{j\beta}^{\dagger} \right >.
\end{equation}
The four-site unit cell of the checkerboard lattice, as depicted in Fig.\ref{fig:cuc},
has twelve of these $Q_{ij}$ fields. When $S$ becomes large, the dynamics of $\mathcal{S}$
leads to the condensation of the $b_i^{\alpha}$ bosons and one obtains a nonzero
value of
\begin{equation}
\left < b^{\alpha}_i \right > = x^{\alpha}_i.
\end{equation}
Since the large-$N$ limit of $\mathcal{S}$
is taken for a fixed value of $\kappa = n_b/N = 2S$, depending on the ratio
$J_1/J_2$ and the value of $\kappa$, the ground state of $\mathcal{S}$ at $T=0$
can either break the global Sp($N$) symmetry and posses magnetic LRO
or be Sp($N$) invariant with only SRO; thus the ground state energy should be
optimized with respect to variations in $\left<Q_{ij}\right>$ and
$x_i^{\alpha}$ for different
values of $J_2/J_1$ and $\kappa$. We also find that each saddle point may be described
by a purely real $\left<Q_{ij}\right>$.
The resulting phase diagram is shown in Fig.\ref{fig:pd}.
We describe various magnetically ordered ($x^{\alpha}_i \not= 0$)
and paramagnetic ($x^{\alpha}_i = 0$) phases as follows.

\subsection{Magnetically ordered phases}

\subsubsection{N\'eel $(\pi,\pi)$ LRO state}
This is the long-range-ordered state in which $\langle\mathbf{S}_i\rangle$
is collinearly polarized in opposite directions on two sublattices.
In the appropriate gauge, the expectation values of link variables are nonzero
and equal on the horizontal (Q) and vertical (P) links, while the values on
the diagonal links (R) are zero.

\subsubsection{$(\pi, 0)$ and $(0, \pi)$ LRO states}
This magnetically ordered phase is characterized by nonzero values of
$\langle Q_{ij}\rangle$ on the horizontal and diagonal links for the
$(\pi, 0)$ LRO state or on the vertical and diagonal links for the
$(0, \pi)$ LRO state. There are two gauge-nonequivalent
choices for the values of $R_1$, $R_2$, $R_3$ and $R_4$. One state has
$R_1=R_2=R_3=R_4$ with $P = 0$ and $Q \not= 0$, and the other has
$R_1=R_3=-R$ and $R_2=R_4=R$ with $P \not= 0$ and $Q = 0$.
These two states are interchanged under a $\pi/2$ rotation.
Moreover, these states can be understood as diagonally coupled horizontal
or vertical Heisenberg antiferromagnetic chains.

\subsubsection{Helical $(\pi,q)$ and $(q,\pi)$ LRO states}
This helically-ordered phase is characterized by nonzero values of
$\langle Q_{ij}\rangle$
on the horizontal, vertical and diagonal links. Once again, our results
indicate that there are two gauge-nonequivalent choices for the values of
$R_1$, $R_2$, $R_3$ and $R_4$. One state has $R_1=R_2=R_3=R_4$ with $Q>P$
and the other has $R_1=R_3=-R$ and $R_2=R_4=R$ with $P>Q$.
These two states are interchanged under a $\pi/2$ rotation and correspond
to spirals ordered in the horizontal or vertical directions.
Notice that this is a long-range incommensurate spin order and the
spin structure factor peaks at the incommensurate wavevector $(\pi, q)$
or $(q,\pi)$. Here $q$ varies continuously as $J_2/J_1$ changes.

\subsection{Paramagnetic phases}

\subsubsection{Decoupled chains}
As its name suggests, this phase is a grid of decoupled Heisenberg
antiferromagnetic chains. In the large-$N$ limit, it corresponds to the
solution where the $\langle Q_{ij} \rangle$ are nonzero only on the
diagonal links.

\subsubsection{$(\pi,\pi)$ SRO state}
This state is obtained by quantum disordering the N\'eel state
(it occurs at $\kappa < 0.4$). The expectation values of $Q_{ij}$
have the same structure as those of the N\'eel state. As described
in previous works, the quantum fluctuations in this phase can be
understood via a compact U(1) gauge theory.\cite{sachdev92}
This theory is always
confining, thus the $b_i^{\alpha}$ bosons bind to form a $S=1$
quasiparticle above a spin gap.\cite{sachdev92}
At finite $N$, gauge-theoretic consideration of singular fluctuations
lead to the plaquette order (see section III).

\subsubsection{$(\pi, 0)$ and $(0, \pi)$ SRO state}
This is the quantum-disordered state of the $(\pi, 0)$ or
$(0, \pi)$ LRO state and the expectation values of $Q_{ij}$ have a
similar structure as those of its ordered counterpart. Even though
this state has a gap to all spin excitations, the symmetry of
$\pi/2$ rotations between the vertical and horizontal directions
is broken. Again, the gauge-theoretic consideration at finite $N$
leads to the staggered spin-Peierls phase (see section III).

\subsubsection{$(\pi,q)$ and $(q,\pi)$ SRO state}
This state is obtained by quantum disordering the $(\pi, q)$
or $(q, \pi)$ LRO state and the expectation values of $Q_{ij}$ have
a similar structure as those of its ordered counterpart.
As discussed in previous works, nonzero expectation values of
the diagonal link fields leads to the presence of the ``charge''-2
Higgs scalar coupled to a compact U(1) gauge field.\cite{sachdev92,marston01}
As a result, this phase corresponds to the deconfined phase of
the $Z_2$ gauge theory and supports deconfined spinons with
$S=1/2$ above the spin gap.\cite{z2,balents01,nayak01,demler02}
Again, even though this state has a gap to all spin excitations,
the symmetry of $\pi/2$ rotations between the vertical and
horizontal directions is broken.
This phase is topologically ordered and it leads to an
additional fourfold degeneracy of the ground
state on a torus.\cite{z2,balents01,nayak01,demler02}

\section{Fluctuation effect and emergence of
translational-symmetry-breaking phases}

In this section, we will discuss the influence of the fluctuations
about the mean-field on the spin-singlet sector of the $(\pi,\pi)$
and $(\pi,0)$ SRO phases. It has been known that the fluctuations
at finite-$N$ can be understood via the compact U(1) gauge
theory.\cite{sachdev92} Regular perturbative corrections in $1/N$
and higher-orders do not qualitatively change the mean-field
ground states at $N \rightarrow \infty$. On the other hand,
singular effects of the hedgehog instantons in the gauge theory
and their Berry phases lead to the qualitative changes, typically
the emergence of various translational-symmetry-breaking phases
\cite{sachdev92}. For simplicity, we will only consider the case
where $2SN$ is an odd integer. Notice that, for the physical SU(2)
case with $N=1$, it corresponds to the half-integer values of $S$.

In the dual formulation, these instanton effects take the form of
a statistical mechanical ``interface'' or ``height'' model. In
most cases this height model has a limiting regime where it can be
mapped onto a quantum dimer model, and this yields an appealing
physical interpretation of the computation. However, the mapping
to the dimer model is {\em not available in all cases\/}, and here
the instanton analysis is a bit more subtle. It turns out that the
$(\pi, 0)$ SRO phase is one of the cases where the dimer model
mapping is absent. Consequently, we will eschew the dimer
interpretation here, and present details of the computation using
the compact U(1) gauge theory.

As has been argued elsewhere \cite{sachdev92,th2002} spin
fluctuations in the paramagnetic phase of collinear
antiferromagnets are described by the following compact U(1)
lattice gauge theory
\begin{eqnarray}
\mathcal{Z}_A &=& \prod_{j\mu} \int_0^{2 \pi} \frac{d A_{j \mu}}{2
\pi} \exp \left(  \sum_{\Box} \frac{1}{K_a}
\cos\left(\epsilon_{\mu\nu\lambda}\Delta_{\nu} A_{j \lambda}
\right) \right. \nonumber \\
&~&~~~~~~~~~~~~~~\left.- i \sum_j \eta_j A_{j\tau} \right),
\label{f5}
\end{eqnarray}
Here $j$ denotes the sites of cubic lattice in discretized
spacetime, the subscript $a$ the sites of the dual cubic lattice,
$\mu$, $\nu$, $\lambda$ extend over $x,y,\tau$, $\Delta_{\mu}$ is
the discrete lattice derivative in the $\mu$ direction, $A_{j
\mu}$ is the compact U(1) gauge field, and the coupling $K_{a}$
can take distinct values of different dual lattice sites as
determined by the symmetry of the underlying lattice. The fixed
field $\eta_j$ is the key quantity which distinguishes different
SRO phases, and encapsulates the staggering of the spins in the
local collinear order. It has the values
\begin{equation}
\eta_j = \left\{ \begin{array}{cc} (-1)^{j_x + j_y} &~\mbox{~~for
$(\pi,\pi)$ SRO} \\
(-1)^{j_x} &~\mbox{~~for $(\pi,0)$ SRO} \end{array} \right.
\label{etaj}
\end{equation}
Note that $\eta_j$ is independent of the $\tau$ co-ordinate.

We now proceed with a duality mapping of (\ref{f5}). For the
$(\pi,\pi)$ SRO phase, the procedure is essentially identical to
that in Ref.~\onlinecite{th2002}, and it leads to a dual interface
model identical to that found for the $(\pi,\pi)$ SRO phase of the
Shastry-Sutherland lattice \cite{marston01}. From these results we
obtain the plaquette-ordered state as shown in Fig. 4(a). Similar
conclusions were also reached in Ref~\onlinecite{sondhi01}.

\begin{figure}
\includegraphics[width=8cm]{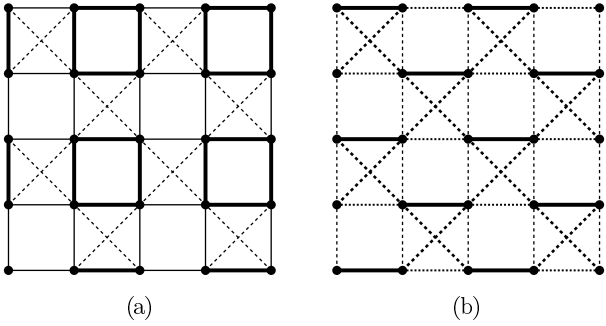}
\caption{\label{fig:4} Singular fluctuations at finite $N$ lead
to (a) the plaquette-ordered phase in the $(\pi,\pi)$ SRO state
and (b) the staggered spin-Peierls phase in the $(\pi,0)$ SRO state.}
\end{figure}

We now apply the same procedure to the $(\pi,0)$ SRO phase. The
only difference here is in the value of $\eta_j$ in (\ref{etaj}),
but otherwise the initial steps are the same as before
\cite{th2002}. First, we find an integer valued vector field,
$a_{a\mu}^0$ on the links of the dual lattice such that its curl
is a vector $\eta_j$ pointing in the $\tau$ direction
\begin{equation}
\epsilon_{\mu\nu\lambda} \Delta_{\nu} a_{a\mu}^0 = \eta_j
\delta_{\mu\tau}.
\end{equation}
\begin{figure}
\includegraphics[width=6cm]{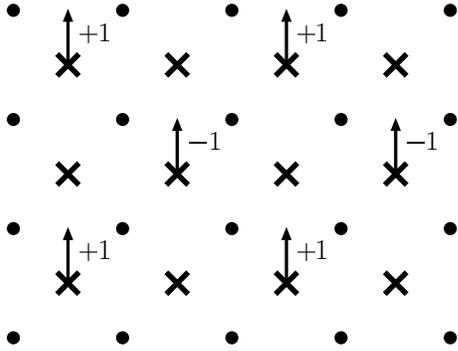}
\caption{\label{fig:a_0_fields} Specification of the non-zero values
of the fixed field $a_{a\mu}^0$. The circles are the sites of the direct
lattice, $j$, while the crosses are the sites of the dual lattice, $a$; 
the latter are also offset by half a lattice spacing in the direction
out of the paper (the $\mu=\tau$ direction). The $a_{a\mu}^0$ are all
zero for $\mu=\tau,x$, while the only non-zero values of $a_{ay}^0$ are
shown above.}
\end{figure}
A convenient choice for $a_{a\mu}^0$ is shown in Fig.~\ref{fig:a_0_fields}.
Then, we write the cosine term in $\mathcal{Z}_A$ in the Villain
periodic Gaussian from, and perform a standard duality
transformation by the Poisson summation method. This maps
$\mathcal{Z}_A$ to a dual interface model
\begin{equation}
\mathcal{Z}_h = \sum_{\{ h_{a} \}} \exp \left( - \sum_{a,\mu}
\frac{K_a}{2}(\Delta_{\mu} h_{a}+ a_{a\mu}^0)^2 \right).
\label{d5}
\end{equation}
Here $h_a$ are the integer valued heights of the interface model.
Next, we decompose $a_{a \mu}^0$ into its curl and divergence free
parts by writing
\begin{equation}
a_{a\mu}^{0} = \Delta_{\mu} \zeta_{a} + \epsilon_{\mu\nu\lambda}
\Delta_{\nu} \mathcal{Y}_{j  \lambda}. \label{XY}
\end{equation}
\begin{figure}
\includegraphics[width=6cm]{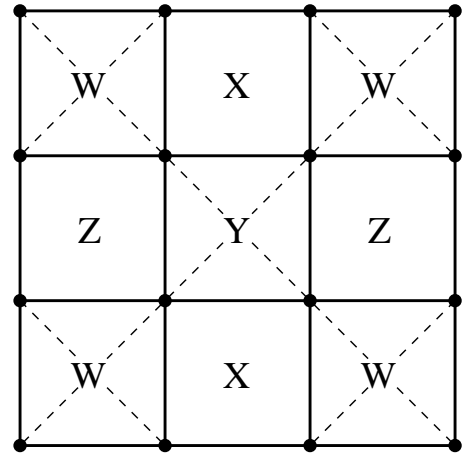}
\caption{\label{fig:duallattice} The four dual sublattices upon
which the height offsets take the values $\zeta_{W}=1/4$,
$\zeta_{X}=1/4$, $\zeta_{Y}=3/4$, and $\zeta_{Z}=3/4$.}
\end{figure}
\begin{figure}
\includegraphics[width=6cm]{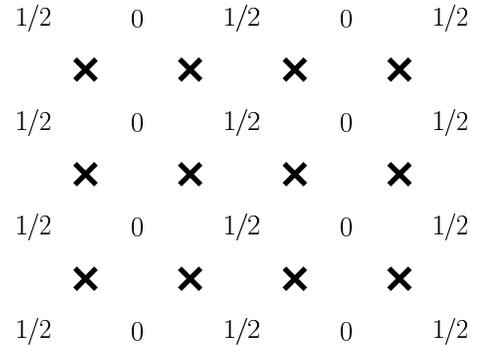}
\caption{\label{fig:y_fields} Specification of the non-zero values
of the fixed field $\mathcal{Y}_{j\mu}$. Only the $\mu=\tau$ components
are non-zero, and these are shown above.}
\end{figure}
The values of the new fields $\zeta_a$ and $\mathcal{Y}_j$ are
shown in Fig.~\ref{fig:duallattice} and Fig.~\ref{fig:y_fields}.
The values of $\zeta_a$ will be particularly important to us, and
these take the values
$\zeta_W=1/4,\zeta_X=1/4,\zeta_Y=3/4,\zeta_Z=3/4$ on the four dual
sublattices $W$, $X$, $Y$, $Z$ shown in Fig.~\ref{fig:duallattice}. Inserting
(\ref{XY}) into (\ref{d5}) we obtain the final form of the
interface model
\begin{equation}
\mathcal{Z}_h = \sum_{\{H_{a}\}} \exp \left ( - \sum_{a}
\frac{K_a^2}{2}\left( \Delta_{\mu} H_{a} \right)^2 \right),
\label{he1}
\end{equation}
where
\begin{equation}
H_{a} \equiv h_{a} + \zeta_{a}. \label{he2}
\end{equation}
Finally, as is usual in the mappings from interface models to
sine-Gordon models, we can `soften' the integer value constraint
on the $h_a$, and replace $H_a$ by a real valued field $\chi_a$
which experiences a cosine potential with minima at positions
satisfying (\ref{he2}). This yields finally a sine Gordon model
for the field $\chi_a (\tau)$ with the action
\begin{eqnarray}
{\cal S}_{\chi} &=& \int  d \tau \Biggl[ \frac{K}{2}
\sum_{\langle ab \rangle} (\chi_a
- \chi_b)^2 \nonumber \\
&+& \sum_a \left\{ \frac{K_{\tau}}{2} (\partial_{\tau} \chi_a)^2 -
y_a \cos(2\pi(\chi_a - \zeta_a)) \right\} \Biggr], \label{schi}
\end{eqnarray}
where the sum over $\langle ab \rangle$ extends over nearest
neighbor sites, and $K$ is the stiffness towards spatial
fluctuations of the interface height. $K_{\tau}$ is the corresponding
stiffness towards time-dependent fluctuations, and, for
simplicity, its value is assumed to be independent of $a$.
The symmetry of the lattice requires that the strength of the periodic
potential take two possible values, $y_a = y_1$ or $y_a = y_2$
depending upon whether the plaquette $a$ has diagonal $J_2$
links across it or not.

The optimal interface configurations can be determined by the
minimization of ${\cal S}_{\chi}$ by a set of time-independent
values of $\chi_W$, $\chi_X$, $\chi_Y$, and $\chi_Z$.
Then, the problem reduces to the minimization of the following
energy as a function of four real variables:
\begin{eqnarray}
E_{\chi} &=& K \Bigl[ (\chi_X - \chi_W)^2 + (\chi_W-\chi_Y)^2
\nonumber
\\ &~&~~+
(\chi_Y-\chi_Z)^2 + (\chi_Z-\chi_X)^2 \Bigr] \nonumber \\
&~&~~- y_1 \Bigl[ \sin(2 \pi \chi_W) -\sin(2 \pi \chi_Y)
\Bigr] \nonumber \\ &~&~~- y_2 \Bigl[ \sin(2 \pi \chi_X) -
\sin(2 \pi\chi_Z) \Bigr]
\end{eqnarray}
Our analysis here is valid only for small $y_1$ and $y_2$,
so we will analytically determine the minima in power
series in $y_{1,2}$. Let us define
\begin{eqnarray}
\chi_W &=& \chi_1 + \chi_2 + \chi_3 \nonumber \\
\chi_X &=& \chi_1 - \chi_2 + \chi_4 \nonumber \\
\chi_Y &=& \chi_1 + \chi_2 - \chi_3 \nonumber \\
\chi_Z &=& \chi_1 - \chi_2 - \chi_4. \label{allchi}
\end{eqnarray}
At the saddle points of $E_{\chi}$, we obtain
\begin{eqnarray}
\chi_2 &=& \frac{\pi^3 (y_2^2-y_1^2)}{16 K^2} \sin(4 \pi \chi_1)
+ {\cal O} (y_{1,2}^4 ) \nonumber \\
\chi_3 &=& \frac{\pi y_1}{2 K} \cos(2 \pi \chi_1) + {\cal O} (y_{1,2}^3) \nonumber \\
\chi_4 &=& \frac{\pi y_2}{2 K} \cos(2 \pi \chi_1) + {\cal O}
(y_{1,2}^3). \label{chis}
\end{eqnarray}
The average interface height, $\chi_1$, is determined by the
minimization of
\begin{equation}
E_{\chi} = E_0 + A \cos(4 \pi \chi_1) + B \cos( 8 \pi \chi_1) +
\ldots, \label{cosines}
\end{equation}
where $E_0$ is a constant independent of $\chi_1$,
\begin{eqnarray}
A &=& -\frac{\pi^2}{2K}(y_1^2+y_2^2)+\frac{\pi^6}{3K^3}(y_1^4+y_2^4) \cr
B &=& \frac{11\pi^6}{96K^3}(y_1^4+y_2^4)-\frac{\pi}{16 K^3}(y_1^2 y_2^2),
\label{ab}
\end{eqnarray}
and all omitted terms are of order $y_{1,2}^6$ or higher.
We now have to minimize (\ref{cosines}) to determine $\chi_1$.
Then from (\ref{chis}), we know $\chi_{2,3,4}$, and hence the
configuration of the interface heights. For small $y_1$ and $y_2$,
$A<0$ and $|A|\gg|B|$, therefore the saddle point solution is
given by $\chi_1 = 0, 1/2$. Taking $\chi_1=0$, we get $\chi_2=0$,
$\chi_3=\frac{\pi y_1}{2K}$ and $\chi_4=\frac{\pi y_2}{2K}$.
Subsequently, using (\ref{allchi}), we obtain
\begin{equation}
\chi_W=\frac{\pi y_1}{2K}, \
\chi_X=\frac{\pi y_2}{2K}, \
\chi_Y=-\frac{\pi y_1}{2K}, \
\chi_Z=-\frac{\pi y_2}{2K}
\end{equation}
Following the previous work by Read and Sachdev,\cite{sachdev92}
the static ``electric field'' in the compact U(1) gauge theory on
the links of the lattice of spins is given by
$iE_{\alpha}=g~\sum_{\beta}\epsilon_{\alpha\beta}\Delta_{\beta}\chi$,
where $\alpha,\beta=x,y$. From this equation, we find $iE_x=\pm
g\pi(y_2+y_1)/(2 K)$ depending on whether the electric field
resides in the even ($+$) or odd ($-$) row, and $iE_y=\pm
g\pi(y_2-y_1)/(2K)$ depending on whether the electric field
resides in an odd ($+$) or even ($-$) column. Finally, these
electric fields can be related to the underlying spin correlation
function: this requires determining the analog of the couplings in
(3.17) of Ref.\onlinecite{rsb}. However, the main effect can also
be deduced here from simple symmetry considerations. The values of
$\eta_j$ in (\ref{etaj}) show that we can define an orientation to
every horizontal link of the square lattice (from $\eta_j = +1$ to
$\eta_j=-1$) but not to the vertical links. Consequently, the
uniform horizontal electric field strengthens every link oriented
parallel to it (say) and weakens those antiparallel to. Also,
there is no corresponding modulation in the vertical direction.
Taken together, these considerations show that the static electric
fields determined above lead to the staggered dimerization pattern
shown in Fig.4 (b).

\section{Conclusion and relation to previous works}

We investigated quantum ground states of the Sp($N$)
antiferromagnetic Heisenberg model on the checkerboard lattice,
where the Heisenberg exchange couplings on the horizontal and
vertical links ($J_1$) are in general not the same as those along the
diagonal links ($J_2$). The mean-field phase diagram
of the model was obtained in the $N \rightarrow \infty$ limit
as a function of $J_2/(J_1+J_2)$ and $1/S$.
We found various magnetically-ordered
phases at large values of $S$, and the corresponding
quantum-disordered paramagnetic phases at small values of $S$.
The effect of singular fluctuations about the mean-field states
was examined in the quantum-disordered paramagnetic phases
and it leads to the emergence of various translational-symmetry
breaking phases.

At small values of $S$, when $J_2$ is small or comparable to
$J_1$, the ground state is the plaquette-ordered phase (shown in
Fig.4 (a)) that is obtained by quantum-disordering the N\'eel
ordered state; this is as predicted in a related study on a
lattice with a similar symmetry \cite{marston01}, and agrees with
the studies of the $J_1=J_2$ case by the exact diagonalization and
large-$N$ studies.\cite{palmer01,sondhi01,fouet01} As $J_2$ is
increased, a topologically ordered $Z_2$-spin-liquid
phase\cite{z2} appears in a narrow region of the phase diagram.
This phase is the quantum-disordered phase of the $(\pi,q)$ LRO
phase. If $J_2$ is further increased, the ground state becomes the
staggered spin-Peierls phase (shown in Fig.4 (b)) that is obtained
by quantum-disordering the $(\pi,0)$ LRO phase. The transition
from the plaquette-ordered (or the staggered spin-Peierls) phase
to the $Z_2$-spin-liquid phase should be described by a $Z_2$
gauge
theory.\cite{sachdev92,marston01,sondhi01,z2,balents01,nayak01,demler02}
On the other hand, in the $J_2 \gg J_1$ limit, the mean-field
ground state is the decoupled-chains SRO phase. At finite $N$, a
weak coupling between the decoupled chains will be generated and
this may lead to the ``sliding-Luttinger-liquid'' phase previously
discovered in the study of the $J_1 \not= J_2$ case with $S=1/2$;
Our large-$N$ study cannot capture this physics.

When $S$ is very large, there exist two magnetically ordered phases;
$(\pi,\pi)$ LRO for $J_2 < J_1$ and $(\pi,0)$ LRO for $J_2 > J_1$.
This result is
consistent with the recent large-$S$ study of the $J_1 \not= J_2$
case.\cite{tcherny03} At intermediate values of $S$, however, there
exists an additional magnetically ordered phase, the $(\pi,q)$ LRO
phase, between the $(\pi,\pi)$ and $(\pi,0)$ LRO phases. This
incommensurate ordered phase is the parent state of the
$Z_2$-spin-liquid phase.

Finally, it is intriguing to see the emergence of the spin-liquid-phase
in the planar pyrochlore (albeit in a narrow region of the phase
diagram) in view of the fact that the existence of a three dimensional
spin-liquid-phase was recently proposed in the easy-axis limit of
the antiferromagnetic Heisenberg model on the three dimensional
pyrochlore lattice.\cite{fisher03}

{\bf Acknowledgment}: We thank Aspen Center for Physics where a
part of this work was carried out, for its hospitality. This work
was supported by the NSERC of Canada, Canadian Institute for
Advanced Research, Canada Research Chair Program, Sloan Fellowship
(JSB, CHC, YBK), and the US NSF grant DMR-0098226 (SS).

\end{document}